\documentclass[groupedaddress,prl,twocolumn,floatfix,nobibnotes,superscriptaddress]{revtex4-1}

\usepackage{graphicx}
\usepackage{dcolumn}
\usepackage{bm}
\usepackage{natbib}
\usepackage{amsmath}
\usepackage{amssymb}
\usepackage{mathtools}
\usepackage{nicefrac}
\usepackage{color}
\usepackage{enumerate}
\usepackage{comment}
\usepackage{framed} 
\usepackage{nicefrac}
\usepackage{units}
\usepackage[subpreambles=true]{standalone}
\usepackage{import}

\renewcommand{\vec}[1]{\boldsymbol{#1}} 
\newcommand{\vechat}[1]{\widehat{\vec{#1}}} 
\renewcommand{\tabcolsep}{0.4cm}

\begin{document}
\title{Non-Reciprocal Spin Pumping in Asymmetric Magnetic Trilayers}
\author{Yevgen Pogoryelov}
\affiliation{Department of Physics and Astronomy, Uppsala University, 751 21 Uppsala, Sweden}

\author{Manuel Pereiro}
\affiliation{Department of Physics and Astronomy, Uppsala University, 751 21 Uppsala, Sweden}

\author{Somnath Jana}
\affiliation{Department of Physics and Astronomy, Uppsala University, 751 21 Uppsala, Sweden}

\author{Ankit Kumar}
\affiliation{Department of Engineering Sciences, Uppsala University, 751 21 Uppsala, Sweden}

\author{Serkan Akansel}
\affiliation{Department of Engineering Sciences, Uppsala University, 751 21 Uppsala, Sweden}

\author{Mojtaba Ranjbar}
\affiliation{Department of Physics, University of Gothenburg, 412 96 Gothenburg, Sweden}

\author{Danny Thonig}
\affiliation{Department of Physics and Astronomy, Uppsala University, 751 21 Uppsala, Sweden}

\author{Peter Svedlindh}
\affiliation{Department of Engineering Sciences, Uppsala University, 751 21 Uppsala, Sweden}

\author{Johan {\AA}kerman}
\affiliation{Department of Physics, University of Gothenburg, 412 96 Gothenburg, Sweden}
\affiliation{Materials and Nanophysics, School of ICT, KTH Royal Institute of Technology, 164 00 Kista, Sweden}

\author{Olle Eriksson}
\affiliation{Department of Physics and Astronomy, Uppsala University, 751 21 Uppsala, Sweden}
\affiliation{School of Science and Technology, \"Orebro University, 701 82 \"Orebro, Sweden}

\author{Olof Karis}
\affiliation{Department of Physics and Astronomy, Uppsala University, 751 21 Uppsala, Sweden}

\author{Dario A. Arena}
\email[Corresponding author: ]{darena@usf.edu}
\affiliation{Department of Physics, University of South Florida, Tampa, Florida 33620, USA}

\begin{abstract}
In magnetic trilayer systems, spin pumping is generally addressed as a reciprocal mechanism characterized by one unique spin mixing conductance common to both interfaces. However, this assumption is questionable in cases where different types of interfaces are present in the material. Here, we present a general theory for analyzing spin pumping in cases with more than one unique interface. The theory is applied to analyze layer-resolved ferromagnetic resonance experiments on the trilayer system Ni$_{20}$Fe$_{80}$/Ru/Fe$_{49}$Co$_{49}$V$_2$ where the Ru spacer thickness is varied to tune the indirect exchange coupling. The results show that the spin pumping in trilayer systems with dissimilar magnetic layers is non-reciprocal, with a surprisingly large difference between spin-pumping induced damping of different interfaces. Our findings have importance on dynamics of spintronic devices based on magnetic multilayer materials.
\end{abstract}

\maketitle

\paragraph{Introduction: }Spin transport in thin film heterostructures can generate a rich spectrum of physical effects and its use has great potential for realizing new spintronic functionality and low power operation \cite{Chen2016procieee,Hellman2017}.  Spin currents without an accompanying charge current can be generated from ferromagnets with a temperature gradient via the spin Seebeck effect \cite{Uchida2008} while the spin Hall effect introduces spin currents from  $Z_2$-topological quantum paramagnets into ferro- / anti-ferromagnets \cite{Hirsch1999,Pai2012}.  Such techniques can be used to inject and transport spin currents \cite{mcdonald, Cornelissen2015,Giles2015}, and even realize new phenomena such as formation of Bose-Einstein superfluids from injected spins \cite{Yuan2018}.  Control of spin currents presents a channel to manipulate magnetic materials via spin transfer torque \cite{Slonczewski1996,Berger1996} without application of an external field.

Spin pumping in layered magnetic materials presents an additional way to generate spin currents.  Pure spin currents can be generated in metallic ferromagnetic (FM) / non-magnetic (NM) heterostructures via spin pumping \cite{Tserkovnyak2002b} where spins excited into precession in a FM generate a spin current in the direction transverse to the static spin direction of the FM, and the spin currents propagate diffusively away from the FM / NM interface and into the NM layer. Propagation of spin currents in the NM can lead to non-local effects such as spin accumulation in the NM \cite{sears} and spin-to-charge current conversion via the inverse spin-Hall effect (ISHE) \cite{Saitoh2006}.  Another characteristic signature of spin pumping is the increased Gilbert-like damping in the FM layer \cite{Mizukami2002}, resulting from the additional loss of angular momentum in the precessing FM system from the spins pumped into the NM \cite{Tserkovnyak2002a}.  Spin pumping can also act as a non-local perturbation of a second FM layer when the NM layer thickness is of the order of the spin diffusion length ($\lambda_{sd}$) or thinner \cite{Heinrich:2003os}, thus providing additional modes of controlling the dynamics of magnetic multilayer structures.  

The efficiency of the spin pumping process across a FM / NM interface is typically parameterized by the spin mixing conductance $g^{\uparrow\downarrow}$ which relates the additional damping from spin pumping in thin films to the film thickness and intrinsic properties such as the saturation magnetization and $g$-factor \cite{Tserkovnyak2002a,Tserkovnyak2002b}.  In magnetic trilayer structures (FM1 / NM / FM2), spin pumping is generally considered as a reciprocal process (FM 1 $\xLeftrightarrow{\text{  SP  }}$ FM 2), characterized by a single $g^{\uparrow\downarrow}$ common to both interfaces; this approach works well when FM1 and FM2 are the same material with an equivalent FM / NM interface \cite{Heinrich:2003rr,Yang2016}.   Many spintronic devices rely on layered magnetic structures where FM1 and FM2 are different materials \cite{Wang2018,Okamoto2015,Deac2008,Mohseni2013,Sani2013,McFadyen2006} with different interfaces on each side of the spacer layer, which casts doubt on an analysis that relies on reciprocal spin-pumping.  

We present in this Letter a framework for analyzing spin pumping in cases where the interfaces are nonequivalent. The theoretical model considers FM1 and FM2 layers with different intrinsic parameters (uniaxial and cubic anisotropy, shape anisotropy, saturation magnetization, and inter-layer exchange coupling or IEC). A key feature of the treatment is the separation of the spin-mixing conductance into distinct contributions for the two dissimilar interfaces.  We apply this theory to analyze layer-resolved ferromagnetic resonance (FMR) experiments measured with x-ray detected FMR (X-FMR) from a series of magnetic trilayer samples where the NM spacer thickness is varied to tune the IEC.  The analysis indicates that the spin pumping from FM1 into FM2 is \textit{non-reciprocal} with the spin pumping in the reverse direction, which can have a considerable effect on dynamics of layers in STOs and related spintronic devices.  

\begin{figure}[htp]
\centering
\includegraphics[width=0.7\linewidth]{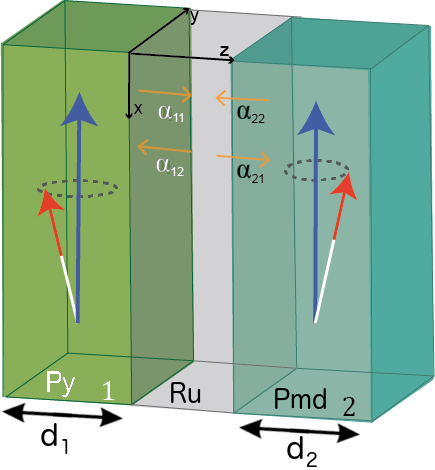}  
\caption{(Color online) Trilayer system composed of Permalloy (1) and Permendur (2) layers separated by a layer of Ruthenium. On the top of the figure the coordinate system used for the  analytical derivations is shown. The red arrow represents the precession of the spin in the $\vechat{y}-\vechat{z}$ plane used in the model while the blue arrow indicates the direction of the external magnetic field. The yellow arrows indicate the static ($\alpha_{ii}$) and dynamic ($\alpha_{ij}$) spin pumping dampings due to the currents flowing in and out of the layers. The thickness of the magnetic layers is represented by $d_1$ and $d_2$.}
\label{fig1}
\end{figure}

\paragraph{Non-Reciprocal Spin Pumping: } We consider a Permalloy (Py - Ni$_{80}$Fe$_{20}$) / NM / Permendur (Pmd - Fe$_{49}$Co$_{49}$V$_2$) magnetically coupled trilayer system with Ru as the NM spacer, denoted as Py /Ru/ Pmd. Here, the first magnetic layer (Py) is labeled in the following by $1$ and the second magnetic layer by $2$ (Pmd), as shown in Fig.~\ref{fig1}. It is assumed that each layer can be represented by a single macrospin $\vec{m}_i$ and $\vec{m}_j$ (macrospin approximation), where $i,j=1,2$. 
The equation of motion of the macrospins can be recast in the form of the Landau-Lifshitz-Gilbert equation:

\begin{equation}
\begin{split}
	\frac{\partial \vec{m}_i}{\partial t}=-\gamma \vec{m}_i\times\vec{H}^i_{\rm eff}&+(\alpha_i^0+\alpha_{ii}^{sp}) \vec{m}_i\times\frac{\partial \vec{m}_i}{\partial t} \\
	&-\alpha^{sp}_{ij} \vec{m}_j\times\frac{\partial \vec{m}_j}{\partial t}
	\label{llg}
\end{split}
\end{equation}
where $\gamma\approx \unit[3.5\cdot10^{-5}]{(GHz\cdot m)/A}$ is the gyromagnetic ratio. Both macrospins ($\vec{m}_i$, $i=1,2$) precess around their effective fields $\vec{H}^i_{\rm eff}=-\nicefrac{\partial \mathcal{H}}{\partial \vec{m}_i}$, where $\mathcal{H}$ is the spin-Hamiltonian. The non-local, dynamic damping, $\alpha^{sp}_{ij}$, accounts for spin-pumping contributions into layer $i$ from layer $j$, while the static damping $\alpha^{sp}_{ii}$ accounts for spin-pumping out of layer $i$. The intrinsic damping of layer $i$ is given by $\alpha_i^0$. 

The coupled trilayer system in the coordinate system indicated in Fig.~\ref{fig1} is described by the following Hamiltonian $\mathcal{H}$:

\begin{align}
\label{hamiltonian}
	\mathcal{H}=&-\frac{A_{ex}}{\mu_0 d_i} \frac{\vec{m}_i\cdot\vec{m}_j}{|\vec{m}_i|| \vec{m}_j|}-\frac{B_{ex}}{\mu_0 d_i} \frac{(\vec{m}_i \cdot\vec{m}_j)^2}{|\vec{m}_i|^2| \vec{m}_j|^2}+\frac{1}{2}\bar{\bar{N_i}} \vec{m}^2_i\nonumber \\
	&-H_0\vechat{x}\cdot\vec{m}_i+\frac{1}{2}\delta_i \vechat{x}\cdot \vec{m}_i -h e^{-i \omega t}\vechat{y}\cdot\vec{m}_i\nonumber \\
	&+\frac{K_i^u}{\mu_0|\vec{m}_i|^2}\left(\vec{m}_i\cdot \vechat{e}_u\right)^2 
	+\frac{K_i^c}{\mu_0|\vec{m}_i|^4}[(\vec{m}_i \cdot\vechat{x})^2 (\vec{m}_i\cdot \vechat{y})^2\nonumber \\
	&+(\vec{m}_i \cdot\vechat{y})^2 (\vec{m}_i\cdot \vechat{z})^2
	+(\vec{m}_i \cdot\vechat{z})^2 (\vec{m}_i\cdot \vechat{x})^2]
\end{align}
where the first ($A_{ex}$) and second ($B_{ex}$) term represent the bilinear- and biquadratic exchange energy, respectively. The next term is the demagnetization energy. In a thin film with $x\gg z$ and $y\gg z$ and uniaxial anisotropy along the $\hat{x}$ axis, the demagnetization tensor $\bar{\bar{N_i}}$ is  non-zero only for $N_i^{xx}=\nicefrac{4 K^u_i}{\mu_0 {m_i^x}^2 }\left(1+2\cos 2\theta_i\right)\sin^2\theta_i$ and $N_i^{zz}=1$. $N_i^{xx}$ represents the contribution to the demagnetization field induced by the in-plane uniaxial anisotropy while $\theta_i$ is the angle between the uniaxial easy axis and the magnetization at layer $i$. The angle dependence of $N_i^{xx}$ was taken from Ref.~\cite{osamu}. The fourth term ($H_0$) represents the energy of a static external magnetic field along the $\vechat{x}$ direction while the following term is the magnetic dipolar field where $\delta_i$ is a term that depends on the structural parameters of the layer as shown in Ref.~\cite{tsymbal}. We consider here only the influence of the field along the $\vechat{x}$ direction since the dipolar field at a distant point (0,0,z) has only x-component for the field. The next term is the microwave field $h e^{-i\omega t}$ oscillating at a frequency $\omega$ along the $\vechat{y}$ direction in the experimental setup. Finally, the last two terms represent the uniaxial (with the easy axis $\vec{e}_u$) and cubic magnetocrystalline anisotropy energy, respectively. The uniaxial anisotropy field lies in the x-y plane. Then, the dependence on $\theta_i$ angle is $\left(\vec{m}_i\cdot  \vechat{e}_u\right)\vechat{e}_u=-2\cos\theta_i(\cos\theta_i,\sin\theta_i,0)$. The parameter $d_i$ indicates the thickness of the ferromagnetic layer while $\mu_0\approx 4\pi\cdot10^{-7} J/(m\cdot A^2)$ is the vacuum permeability. 

\begin{figure*}[htp]
\centering
	\includegraphics[width=1.00\linewidth]{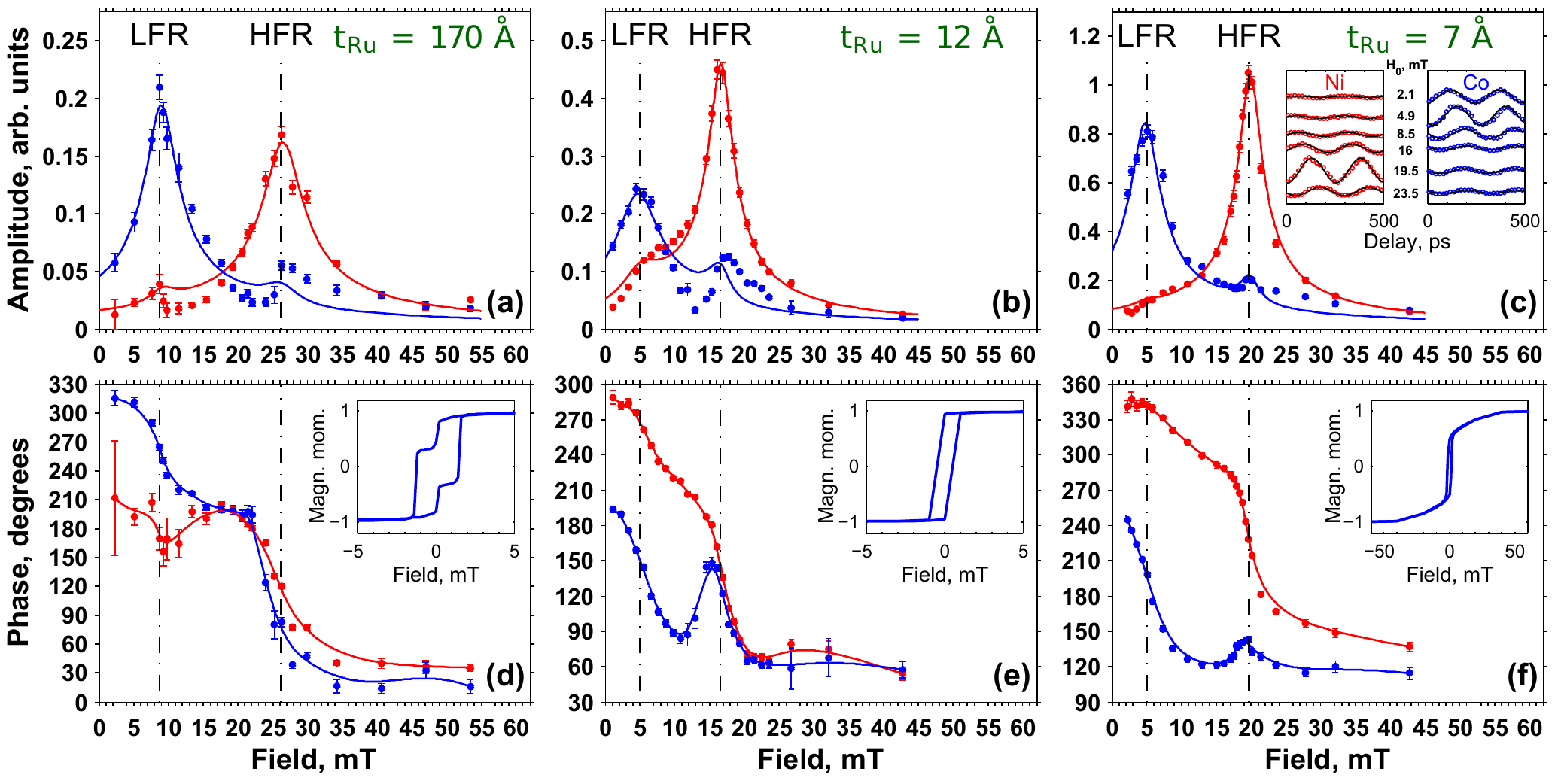}
	\caption{(color online) Non-reciprocal spin pumping theory and X-FMR measurements.  Theoretical calculations (solid lines) and measured X-FMR data (data points) shown for three different Ru spacer layer thicknesses: 170 \AA (a,d) 12 \AA (b,e) and 7 \AA (c,f).  Data and simulations for the Pmd layer are shown in blue while the Py is in red.  Top row shows the amplitude of the resonant response while the bottom row shows the phase of the X-FMR response. See text for details. Insets in (d,e,f) show the static magnetometry data.}
\label{fig2}
\end{figure*}

Since the moments rotate around the external magnetic field, the condition $m^x\gg m^y, m^z $ is fulfilled and, consequently, $\frac{d m_i^x}{dt}=0$ and $m_i^y(t)=m_i^y\exp[-i(\phi_i+\omega_i t)]$, $m_i^z(t)=m_i^z\exp[-i(\phi^\prime_i+\omega_i t)]$ with $\phi_i^\prime=\phi_i+\pi/2$ and $ \omega=\omega_1=\omega_2$. The phase $\phi_i$ of magnetic layer $i$ is measured by X-FMR experiments. Assuming that the angle of precession of the macrospin is relatively small ($m^x\gg m^y, m^z $) and also that $h\ll H_0$, it is a good approximation to linearize the equations of motion. Thus, only terms linear in $h$, $ m^y$ and  $m^z$ are retained. Moreover, we also assume that the easy axis of the  uniaxial anisotropy is along $\vechat{x}$ so that $\vechat{e}_u=(1,0,0)$. By inserting Eq.~(\ref{hamiltonian}) into Eq.~(\ref{llg}) through the definition of the effective field, the linearized coupled equations of motion for both magnetic layers are given by:
\begin{equation}
\bar{\bar{\chi}}
	\begin{pmatrix}
		h\\
		0\\
		h\\
		0\\
	\end{pmatrix}
	=
	\begin{pmatrix}
		m^y_1\\
		m^z_1\\
		m^y_2\\
		m^z_2\\
	\end{pmatrix}
		\label{responce-mom}
	\end{equation} 
where the magnetic susceptibility matrix, $\bar{\bar{\chi}}$ is derived in the supplemental material. Hereafter, the dimensionless intrinsic damping parameter is defined as $\eta_i^0=m_i^x \alpha_i^0$. The spin pumping damping parameter out of layer $i$ is $\eta^{sp}_{ii}= m_i^x \alpha^{sp}_{ii}$, while the dimensionless spin pumping damping parameter into layer $i$ from layer $j$ is defined as $\eta^{sp}_{ij}=m_j^x \alpha^{sp}_{ij}$. The amplitude of the macrospin precession shown in Fig.~\ref{fig2} (a) -- (c) is calculated from the four-index susceptibility matrix \cite{footnote} of the system, $\bar{\bar{\chi}}$, as:
\begin{align}
	\psi_1&=\sqrt{[\mathfrak{R}(\chi_{21}+\chi_{23})]^2+[\mathfrak{I}(\chi_{21}+\chi_{23})]^2} \label{susceptibility2} \\
	\psi_2&=\sqrt{[\mathfrak{R}(\chi_{41}+\chi_{43})]^2+[\mathfrak{I}(\chi_{41}+\chi_{43})]^2}.
	\label{susceptibility1}
\end{align}

\paragraph{X-FMR Measurements: }
X-ray detected ferromagnetic resonance \cite{Bailey2004, Martin:2009, Marcham2011}, or X-FMR, is the ideal technique to investigate non-reciprocal spin pumping. Using X-FMR technique it is feasible to measure the full complex susceptibility ($\chi '$ and $\chi ''$) \cite{Arena_RSI2009} resolved to individual elements \cite{Warnicke2015PRB} and hence distinct magnetic layers \cite{Arena2006,Guan:2006ul,Bailey_NatComm2013}. X-FMR was previously used to spin pumping and the influence of spin currents \cite{Li2016,Baker2015,Figueroa2016}. For the investigation of non-reciprocal spin pumping, the measured motion of the individual layers via X-FMR can be compared directly with the equation of motion [Eq.~(\ref{llg})] for FM1 and FM2.

We use the X-FMR technique to study a series of Py / Ru / Pmd magnetic trilayer film structures (see Supplemental Material for details of the sample structure and preparation). The Ni in Py and Co in Pmd provide the elemental contrast that permits X-FMR to resolve the dynamics in the individual FM layers while Ru produces a strong IEC that can be tuned from favoring parallel or anti-parallel ground state coupling as a function of the NM spacer thickness \cite{Parkin1991}.

In X-FMR experiments we perform time delay scans, which are equivalent to varying the phase between the sinusoidal RF signal and the arrival of the x-ray photons.  The inset to Fig.~\ref{fig2} (c) presents a subset of these delay scans; for further details, refer to  \cite{Arena_RSI2009}.  The simple sinusoidal waveforms of the delay scans allow us to extract the amplitude and phase of the precessional motion and these parameters are shown in Fig.~\ref{fig2} (a) -- (f) as discrete points.  

Starting with the amplitude data in Fig.~\ref{fig2} (a) for the $t_{Ru}$~=~170~{\AA} sample (no bilinear or biquadratic IEC), we observe two resonances at $\sim$27 mT and $\sim$9 mT.   While the two resonances are associated predominantly with the Pmd (low field) and Py (high field) layers, the coupled nature of the FM layer dynamics precludes assignment of the combined resonance to a specific layer and we refer to these as a low field and high field resonance (LFR or HFR).  The other samples also present a LFR and HFR.  However, it is clear in examining the amplitude data that at any resonant field, a particular FM layer does not respond independently and that the other layer also exhibits a distinct response, albeit weaker.  For example, in  Fig.~\ref{fig2} (a), where the Py (Pmd) layer has a maximum at the HFR (LFR), and a clear increase in the amplitude is apparent at the LFR (HFR).

The phase data also reveal a non-trivial, coupled, response of the oscillation phase as the magnetic field is swept through the resonance.  For independent layers, the phase of FM1 would change by 180$^\circ$ when passing through the resonant field while the other layer would remain unchanged.  For layers that are coupled via IEC, dipolar effects, and in particular spin pumping, the phases of the dynamic response would change according to Eq.~(\ref{responce-mom}).  In the sample with $t_{Ru}$~=~170~{\AA} (Fig.~\ref{fig2} (d)), and weakest coupling between Py and Pmd, the phase of the Py layer changes by $\sim$ 180$^\circ$ through the HFR.  However, the Pmd layer undergoes approximately the same phase shift through the HFR and then experiences an additional phase shift of $\sim$120$^\circ$ through the LFR.

The deviations from the independent layer expectation  are even more dramatic for the more strongly coupled samples ($t_{Ru}$~=~12~{\AA} and 7\AA).  In the FM-coupled $t_{Ru}$~=~12~{\AA} sample the two phases are essentially equal from high field down through the HFR at  $\sim$16 mT.  As the field is reduced, the Pmd phase drops dramatically away from the Py layer before rising again as the LFR is approached.  For the $t_{Ru}$~=~7~{\AA} sample with coupling between FM and AFM, the transition towards AFM coupling serves to keep the phases of the two layers separate.  Although there is an uptick in the response of the Pmd layer at the HFR field at 20 mT, for the most part the two layers maintain a large phase separation ($>$100$^\circ$) between the HFR and the cut off at zero field.

\paragraph{Analysis and Discussion: }
We begin our analysis with the phase data in Fig.~\ref{fig2} (d), (e) and (f).  The experimental phases are first fit with a high order B-spline and we use these interpolated values for the phase, together with the other parameters for magnetization, anisotropy and interlayer exchange, to calculate the amplitude response based on Eqs.~ (\ref{susceptibility2})-(\ref{susceptibility1}).  The results are presented as the solid lines in Fig.~\ref{fig2} (a), (b), and (c).  Apart from the phase data, all sample parameters used in the amplitude calculations (\textit{i.e.} magnetization, anisotropy constants, layer thickness, interlayer exchange parameters, \textit{etc.}) were obtained from independent measurements (see supplemental information).  

Generally, the model is in good agreement with the experimental data in Fig.~\ref{fig2} across the whole applied field range.  The peak positions and amplitudes are well-reproduced.  The model captures not only the main resonance associated to one of the magnetic layers, but also the weaker response connected to the second layer. For example, in the most strongly coupled tri-layer sample (Fig.~\ref{fig2} (c), $t_{Ru}$ = 7 \AA),  the model reproduces the increased amplitude in the response of the Pmd layer at the HFR. For the ferromagnetically coupled sample (Fig.~\ref{fig2} (b), $t_{Ru}$ = 12 \AA), the model accurately predicts the reduced amplitude of the Pmd response at 5 mT (blue) in comparison with the main Py resonance (red) at 16 mT. The model also captures the experimental data, both in approximate field and relative amplitude, in the Py response at 5 mT and the Pmd response at 16 mT.  The sample with the thickest Ru spacer (Fig.~\ref{fig2} (a), $t_{Ru}$ = 170 \AA), where interlayer exchange and dipolar coupling effects are negligible, is particularly interesting.  For completely decoupled layers, we would expect that the amplitude of the two resonances can be described by smooth, symmetric Lorentzian functions.  However, the experimental data show an increase of the Pmd (Py) amplitude at the resonant field of the Py (Pmd) layer, and this effect is clearly reproduced by the theory.

By using the measured phase response, along with the independently derived material parameters, the model described by Eqs.~(\ref{llg}-\ref{susceptibility1}) provides estimates of the precessional damping and the contributions from spin pumping (Table~\ref{table1}).  We focus our discussion on the spin mixing conductance of layer $i$, ($g_i^{\uparrow\downarrow}$) related to the dimensionless spin pumping parameter $\eta_{ij}$. In particular it has been shown that the real part of $g_i^{\uparrow\downarrow}$ relates to $\eta_{ij}$  as \cite{Durrant2017}
\begin{equation}
	\mathfrak{R}(g_i^{\uparrow\downarrow})=\frac{8\pi m_j^x d_j \eta_{ji}}{g_j \mu_B},\:\text{where}\:i\neq j.
\end{equation}
Here, g$_j$ is the electron g-factor and $\mu_B$ is the Bohr magneton. The main finding of this work is that the contributions of the two interfaces (Py / Ru or Ru / Pmd) to the spin mixing conductance is not reciprocal. For all the samples studied, we observe that the real part of the spin-mixing conductance from the Py layer, that influences the dynamics of the Pmd layer, is an order of magnitude larger than the reversed spin-mixing conductance, which is close to values reported in the literature ~\cite{Durrant2017}.

\begin{table}[ht]
\caption{Estimated spin pumping induced damping parameters and real part of the spin mixing conductance in cm$^{-2}$ for samples with $t_{Ru}$~=~7~{\AA}, 12~{\AA} and 170~{\AA}.}
\renewcommand\tabcolsep{5.7pt}
\begin{tabular}{lcccccc}
\hline
\hline
$t_{Ru}$ & {$\eta^{sp}_{11}$}  & $\eta^{sp}_{22}$ & {$\eta^{sp}_{12}$} & $\eta^{sp}_{21}$ & $\mathfrak{R}$(g$_1^{\uparrow\downarrow}$) &$\mathfrak{R}$(g$_2^{\uparrow\downarrow}$)\\
\AA   &$10^{-3}$&$10^{-3}$&$10^{-3}$&$10^{-3}$&$10^{15}$&$10^{15}$ \\
\hline
7 & 2.13 & 9.58 & 0.89 & 25.91&50& 0.69\\
12 &2.90 &16.08& 8.46& 30.00& 58& 6.5\\ 
170 &9.73 &9.01& 3.00&31.00&60 &2.3 \\                                          
\hline
\hline
\end{tabular}
\label{table1}
\end{table}

In the standard picture of spin pumping, a magnetic layer excited into precession drives a diffusive spin current in the direction transverse to the FM1 / NM interface.  The spin current incident upon the NM layer leads to spin accumulation in the NM near the interface and generates a flow of spin current back to the FM1.  The spin mixing conductance parametrizes the balance of the initial spin current (FM1 $\rightarrow$ NM) and the backflow into the magnetic layer. For NM layers that are thin compared to the spin diffusion length in the NM ($\sim$15 nm in Ru \cite{Behera2015}) the spin current driven across the NM layer transports angular momentum across the NM / FM2 interface (that has its own characteristic spin mixing conductance), thereby influencing the dynamics of FM2.  The two spin mixing conductances are often assumed to be equivalent \cite{Durrant2017}.  However, the multilayer spin pumping theory presented above together with the X-FMR results clearly indicate that spin pumping is non-reciprocal in systems with nonequivalent interfaces\cite{Zwierzycki2005,Xia2002,Liu2014}, and that the asymmetry between the two interfaces can be substantial. 

Interface spin transport governs a variety of phenomena such as spin injection into semiconductors, topological insulators, and two-dimensional materials, or the generation of pure spin currents via the spin Hall effect \cite{Durrenfeld2015,Roschewsky2016,Zhang2015b,Pai2012} and the related issue of determination of spin Hall angles \cite{Zhang2015b}.  Spin pumping presents another method for manipulating magnetization across an interface, allowing even for non-local effects. Our analysis extends spin pumping theory towards more general magnetic multilayer structures, which may also have distinct layer anisotropies, magnetization and tunable IEC.  These effects can influence the dynamics of individual layers and when these issues are assessed independently, the asymmetry of the spin pumping contributions is revealed.  To our knowledge, this is the first report of non-reciprocal spin pumping in magnetic trilayers with dissimilar interfaces, and our findings may open new possibilities in spintronics technology.  Earlier first principles calculations of spin pumping indicate that the matching of states in the NM with spin-resolved propagating states in the FM layer, greatly affects spin transmission and reflection across the interface \cite{Zwierzycki2005,Xia2002}.  Our analysis supports this viewpoint and also demonstrates that measurable differences in interfacial spin pumping can uniquely be revealed in X-FMR. Of particular interest would be experiments of quasi-epitaxial FM1 / NM / FM2 trilayer structures with well-controlled interfaces which could be compared directly with first-principles calculations of spin-dependent band structure across the two interfaces.  

The authors gratefully acknowledge the support of the Knut and Alice Wallenberg foundation, the Swedish research council (VR) under contracts 2016-04524,and 2013-08316. This research used resources of the Advanced Photon Source operated for the DOE Office of Science by Argonne National Laboratory under Contract No. DE-AC02-06CH11357.

\bibliographystyle{apsrev4-1}

\begin{thebibliography}{42}
\bibitem{footnote} Notice that in Eqs.~(\ref{susceptibility2})-(\ref{susceptibility1}) the sub-indexes of the susceptibility refer to its matrix elements.
\end{thebibliography}

\clearpage

\setcounter{table}{0}
\makeatletter 
\renewcommand{\thetable}{SI\@arabic\c@table}
\makeatother

\begin{widetext}
\begin{center}
\Large\textbf{Non-Reciprocal Spin Pumping in Asymmetric Magnetic Trilayers \\ ---Supplemental Material---}
\end{center}

\begin{center}
Yevgen Pogoryelov,$^1$ Manuel Pereiro,$^1$ Somnath Jana,$^1$ Ankit Kumar,$^2$ Serkan Akansel,$^2$ Mojtaba Ranjbar,$^3$ Danny Thonig,$^1$ Olle Eriksson,$^{1,4}$ Peter Svedlindh,$^2$ Johan {\AA}kerman,$^{3,5}$ Olof Karis,$^1$ and Dario A. Arena$^6$\\
\it{$^1$Department of Physics and Astronomy, Uppsala University, 751 21 Uppsala, Sweden\\
$^2$Department of Engineering Sciences, Uppsala University, 751 21 Uppsala, Sweden\\
$^3$Department of Physics, University of Gothenburg, 412 96 Gothenburg, Sweden\\
$^4$School of Science and Technology, \"Orebro University, 70182 \"Orebro, Sweden\\
$^5$Materials and Nanophysics, School of ICT, KTH Royal Institute of Technology, 164 00 Kista, Sweden\\
$^6$Department of Physics, University of South Florida, Tampa, Florida 33620, USA}
\end{center}

\section{Sample preparation and characterization}
The film samples were fabricated at room temperature using dc magnetron sputtering  (base pressure of $5\times10^{8}$ Torr) with the following structure: substrate/Ta(30~{\AA})/Py(80~\AA)/Ru($t_{Ru}$)/Pmd(80~\AA) /Ta(30~\AA). Here $t_{Ru}$ varies between 7 -- 170~\AA. Single films of Pmd and Py with the same seed and cap layers were also fabricated for control measurements. The Ru spacer layer was deposited at low sputtering rate (0.4 {\AA}/s) and low Ar gas pressure (3 mTorr) for optimal uniformity and interface smoothness. Composition and thickness of the films was verified with the Rutherford Back Scattering (RBS) technique. X-ray reflectivity (XRR) measurements were also performed to check the quality of the layers and interfaces. Each sample was fabricated simultaneously on an oxidized Si-substrate for magnetometry and structural measurements and on a 100~nm thick Si$_{3}$N$_{4}$ membranes for X-FMR. To minimize the number of free parameters in Eq.~(\ref{hamiltonian}), we conducted a series of static magnetometry and FMR measurements on the samples.

\section{Static magnetometry and FMR measurements}
Field hysteresis measurements of single magnetic layers of the control samples provided the saturation magnetization for each magnetic layer: 4$\pi m_{Py}$~=~0.89$\cdot 10^7$ A/m and 4$\pi m_{Pmd}$~=~2.25$\cdot 10^7$ A/m. Magnetization curves for a selected number of samples with $t_{Ru}$ = 7, 12, 170 {\AA} are shown as insets in Fig.~\ref{fig2} (f), (e) and (d) respectively. The $t_{Ru}$~=~7~{\AA} sample shows the behavior typical for a 90$^{\circ}$ coupling between Py and Pmd layers. A Ru thickness of $t_{Ru}$~=~12~{\AA} favors ferromagnetic (FM) coupling between magnetic layers and for a thick Ru spacer  ($t_{Ru}$~=~170~{\AA}) the magnetic layers are de-coupled. Magnetometry results correlate well with the FMR measurements.

We measured the in-plane uniaxial anisotropy constants with angular dependent X-band (9.8 GHz) FMR (rotation about the surface normal) while the interlayer exchange constants ($A_{ex}$ and $B_{ex}$) were determined from in-plane FMR measurements at varying excitation frequencies (2 - 12 GHz). The angular dependent X-band measurements indicate that all samples exhibit a weak uniaxial anisotropy, with the largest anisotropy constant at about $K_{Pmd}^u$~=~2408~$J/m^3$ for Pmd and $K_{Py}^u$~=~184~$J/m^3$ for Py. Cubic anisotropy was found to be negligibly small: $K_{Pmd}^c$~=~179~$J/m^3$ and $K_{Py}^c$~=~10.6~$J/m^3$. Note that we consider the anisotropy constants as independent of the spacer layer thickness, although it may change. All measured parameters are summarized in Table \ref{tableS1}.

\begin{table}[ht]
\caption{Measured physical magnitudes for Permalloy (Py) and Permendur (Pmd) layers. }
\begin{tabular}{lccccc}
\hline
\hline
Layer& {$K_i^u$}  & {$K_i^c$} & $d_i$ & $m_i^x$ & $\omega$\\
         &{($J/m^3$)}& {($J/m^3$)} & (m)& ($A/m$)&  (GHz)\\
\hline
Py ($i=1$) &184.142 & 10.624&$8\cdot10^{-9}$ & $(0.89 \cdot 10^7)/(4 \pi)$&3.96\\
Pmd ($i=2$) &2408.213 & 179.049& $8\cdot10^{-9}$& $(2.25 \cdot 10^7)/(4 \pi)$&3.96\\                       
\hline
\hline
\end{tabular}
\label{tableS1}
\end{table}

We determined $A_{ex}$ and $B_{ex}$ from fits of the resonant field vs. frequency as outlined in Ref. \cite{Wei_TriLayer_FMR_JAP2013}. The sample with the thickest NM spacer layer ($t_{Ru}$~=~170~{\AA}) does not present any bilinear or biquadratic coupling, consistent with the M vs. H loops which show the switching of the individual layers (See inset in Fig. 2 (d)).  As the Ru thickness decreases, interlayer-exchange coupling begins to correlate the switching of the two layers.  The $t_{Ru}$~=~12~{\AA} sample shows FM coupling between magnetic layers with only a bilinear type of coupling present $A_{ex} =\unit[4.5\cdot 10^{-11}]{J/m^2}$; the field hysteresis loops confirm this as only a single switching field is evident (See inset in Fig.~2~(e)). Finally, for the $t_{Ru}$~=~7~{\AA} sample we find the bilinear exchange coupling constant $A_{ex} = 0 $, although there is a large biquadratic coupling parameter $B_{ex} = \unit[-7\cdot 10^{-5}]{J/m^2}$. This indicates that the coupling of the two layers is shifting from FM to AFM, leaving a $\sim$90$^{\circ}$ coupling between the Py and Pmd layers. Exchange constants are summarized in Table \ref{tableS2}.

\begin{table}[ht]
\caption{Estimated exchange constants for samples with $t_{Ru}$~=~7~{\AA}, 12~{\AA} and 170~{\AA}. }
\begin{tabular}{lcc}
\hline
\hline
$t_{Ru}$& $A_{ex}$  & $B_{ex}$  \\
         &{($J/m^2$)}& {($J/m^2$)}  \\
\hline
7 & 0 & $-7\cdot 10^{-5}$ \\
12 &$4.5\cdot10^{-11}$ & 0\\ 
170 &0 &0 \\                                          
\hline
\hline
\end{tabular}
\label{tableS2}

\end{table}

\section{X-FMR Measurements}

X-FMR combines two powerful spectroscopic techniques: x-ray magnetic circular dichroism (XMCD) and FMR \cite{Bailey2004, Martin:2009, Marcham2011}. XMCD provides the elemental contrast to discriminate the dynamics from FM1 or FM2 while FMR measures dynamic parameters such as the effective gyromagnetic ratio and the Gilbert damping constant.  In X-FMR, a microwave field excites precession of the magnetic moments while an external DC magnetic field is swept to tune the response of the magnetic layer through resonance. In our experiments we use a sinusoidal radio frequency (RF) excitation at 3.96 GHz that is synchronized with the bunch repetition frequency (BRF) of x-rays from a synchrotron storage source (BRF = 88 MHz at  Advanced Photon Source, Argonne National Lab), thus ensuring a well-defined phase relationship between the RF excitation and the x-ray bunches.  We use circularly polarized x-rays to take advantage of the XMCD magnetic and elemental contrast and, by tuning the x-ray energy to the Ni or Co $L_3$ edge during the scans, we can distinguish the dynamic response of the Py (FM1) or Pmd (FM2) layer.

\section{Non-Reciprocal Spin Pumping Theory}

The magnetic susceptibility, which is a $4\times 4$ matrix (according to Eq.~\eqref{responce-mom} of the main part of the paper), is defined as:

\begin{equation}
    \bar{\bar{\chi}}=\bar{\bar{\mathcal{A}}}^{-1}
\end{equation}
where the elements of the matrix $\bar{\bar{\mathcal{A}}}$ are:
	\begin{align*}
	a_{11}=&\biggl(\frac{-2 K_1^c}{\mu_0 {m_1^x}^2}-\frac{H_0-(\tau_1(\theta_1)+\delta_1) m_1^x}{m_1^x}-\frac{2 K_1^u}{\mu_0 {m_1^x}^2}\left(\cos\theta_1\sin\theta_1+\cos^2\theta_1\right)+\frac{A_{ex}}{\mu_0 d_1 {m_1^x}^2}\nonumber\\
	&+\frac{2 B_{ex}}{\mu_0 d_1 {m_1^x}^2}+\frac{(\eta_1^0+\eta_{11}^{sp})i\omega}{\gamma m_1^x} \biggr)e^{-i \phi_1},\nonumber\\
	a_{12}=&\frac{-\omega e^{-i \phi_1}}{\gamma m_1^x},\nonumber\\
	a_{13}=&\left(\frac{A_{ex}}{\mu_0 d_1 m_1^x m_2^x}-\frac{\eta_{12}^{sp} i \omega}{\gamma m_1^x} \right)e^{-i \phi_2},\nonumber\\
	a_{14}=&0,\nonumber\\
	a_{21}=& \frac{-\omega e^{-i \phi_1}}{\gamma m_1^x},\nonumber\\
	a_{22}=& \biggl(\frac{-2 K_1^c}{\mu_0 {m_1^x}^2}-1+\frac{H_0-(\tau_1(\theta_1)+\delta_1) m_1^x}{m_1^x}+\frac{2 K_1^u}{\mu_0 {m_1^x}^2}\cos^2\theta_1+\frac{A_{ex}}{\mu_0 d_1 {m_1^x}^2}+\frac{2 B_{ex}}{\mu_0 d_1 {m_1^x}^2}\nonumber \\
	&+\frac{(\eta_1^0+\eta_{11}^{sp})i\omega}{\gamma m_1^x} \biggr)e^{-i \phi_1},\nonumber\\
    a_{23}=& 0,\nonumber\\
    \end{align*}
	\begin{align*}
	a_{24}=& \left(\frac{A_{ex}}{\mu_0 d_1 m_1^x m_2^x}-\frac{\eta_{12}^{sp} i \omega}{\gamma m_1^x} \right)e^{-i \phi_2},\nonumber\\
	a_{31}=&\left(\frac{A_{ex}}{\mu_0 d_2 m_1^x m_2^x}-\frac{\eta_{21}^{sp} i \omega}{\gamma m_2^x} \right)e^{-i \phi_1}, \nonumber\\
	a_{32}=&0, \nonumber\\
	a_{33}=&\biggl(\frac{-2 K_2^c}{\mu_0 {m_2^x}^2}-\frac{H_0-(\tau_2(\theta_2)+\delta_2) m_2^x}{m_2^x}-\frac{2 K_2^u}{\mu_0 {m_2^x}^2}\left(\cos\theta_2\sin\theta_2+\cos^2\theta_2\right)+\frac{A_{ex}}{\mu_0 d_2 {m_2^x}^2}\nonumber\\
	&+\frac{2 B_{ex}}{\mu_0 d_2 {m_2^x}^2}+\frac{(\eta_2^0+\eta_{22}^{sp})i\omega}{\gamma m_2^x} \biggl)e^{-i \phi_2}, \nonumber\\
	a_{34}=& \frac{-\omega e^{-i \phi_2}}{\gamma m_2^x},\nonumber\\
	a_{41}=&0, \nonumber\\
	a_{42}=&\left(\frac{A_{ex}}{\mu_0 d_2 m_1^x m_2^x}-\frac{\eta_{21}^{sp} i \omega}{\gamma m_2^x} \right)e^{-i \phi_1}, \nonumber\\
	a_{43}=&\frac{-\omega e^{-i \phi_2}}{\gamma m_2^x}, \nonumber\\
	a_{44}= &\biggl(\frac{-2 K_2^c}{\mu_0 {m_2^x}^2}-1+\frac{H_0-(\tau_2(\theta_2)+\delta_2) m_2^x}{m_2^x}+\frac{2 K_2^u}{\mu_0 {m_2^x}^2}\cos^2\theta_2+\frac{A_{ex}}{\mu_0 d_2 {m_2^x}^2}+\frac{2 B_{ex}}{\mu_0 d_2 {m_2^x}^2}\nonumber\\
	&+\frac{(\eta_2^0+\eta_{22}^{sp})i\omega}{\gamma m_2^x} \biggr)e^{-i \phi_2}. \nonumber\\
	\nonumber
\end{align*}
Here, we applied 

\begin{align*}
\tau_i\left(\theta_i\right)=\frac{4K_i^u}{\mu_0 (m_i^x)^2}\left(1+2\cos(2\theta_i)\right)\sin^2\theta_i.    
\end{align*}

By using the data collected in Tables~\ref{tableS1}-\ref{tableS2} and the measured phase shown in Fig.~\ref{fig2}, the model described by Eq.~(\ref{responce-mom}) provides the spin-pumping dampings, angles $\theta_i$ and dipolar field prefactors. These data is collected in Tables~\ref{table1} and ~\ref{tableS3}.

\begin{table}[ht]
\caption{Angle between easy-axis and magnetization as well as the layer-dependent dipolar field prefactor for samples with $t_{Ru}$~=~7~{\AA}, 12~{\AA} and 170~{\AA}.}
\begin{tabular}{lcccc}
\hline
\hline
$t_{Ru}$  &{$\theta_1$} &{$\theta_2$}& $\delta_1$&$\delta_2$\\
       &rad&rad& &  \\
\hline
7 &  0.07&-0.73&0.0263 &0.0014\\
12 & 1.38 &0.63&-0.0075 &-0.0018\\ 
170 & 2.94&-0.35& 0.0043&0.0000\\                                          
\hline
\hline
\end{tabular}
\label{tableS3}
\end{table}


\end{widetext}

\end{document}